\documentclass[12pt]{article}
\usepackage[left=2.5cm,top=2.5cm,right=2.5cm, bottom=2.5cm, a4paper]{geometry}
\usepackage{multirow}
\usepackage{enumerate}
\usepackage{bm}
\usepackage{bbm}
\usepackage{booktabs}
\usepackage{natbib}
\usepackage{amssymb}
\usepackage{graphicx}
\usepackage{subfig}
\usepackage{caption}

\begin{document}

%
\title{Bayesian nonparametrics for \\stochastic epidemic models}
\author{Theodore Kypraios\footnote{theodore.kypraios@nottingham.ac.uk} and Philip D. O'Neill\footnote{philip.oneill@nottingham.ac.uk} \\ \\ 
\small{School of Mathematical Sciences, University of Nottingham, UK}}
\date{}

\maketitle

\begin{abstract}
The vast majority of models for the spread of communicable diseases are parametric in
nature and involve underlying assumptions about how the disease spreads through a population.
In this article we consider the use of Bayesian nonparametric approaches to analysing data from
disease outbreaks. Specifically we focus on methods for estimating the infection process
in simple models under the assumption that this process has an explicit time-dependence.
\end{abstract}

%

\section{Introduction}
In this article we describe some recent developments in the field of Bayesian nonparametric inference
for stochastic epidemic models. The topic is itself relatively new; methods for fitting epidemic
models to data are overwhelmingly based on parametric approaches, as described in the other articles
in this edition. To clarify our terminology, we shall refer to {\em models} as either parametric or nonparametric,
whilst references to parametric or nonparametric {\em inference} simply refer to inference for the kind of
model in question, as opposed to the inference methods {\em per se}. We start by motivating the nonparametric
approach, review some classical methods, and then give a brief overview of the remainder of this article.

\subsection{Motivating nonparametrics for epidemic models} Epidemic models, whether stochastic or
deterministic in nature, are almost invariably {\em mechanistic}. This means that such models attempt to describe
the processes which generate observable quantities, usually by making assumptions
about the underlying biological and epidemiological processes involved. For example, suppose we have data consisting
of new cases of a particular disease on a daily basis. The modelling approach involves
defining a model which describes how new cases come about, typically by making assumptions about infection
processes between infected and healthy individuals, and also about the progression of disease within a typical individual.

For stochastic models, such underlying assumptions often involve {\em parametric} models of some kind. Common examples include
(i) assuming a Poisson process to describe the times of contacts between individuals, where the rate parameter
is a model parameter; (ii) assuming that different stages of disease, such as the latent period or infectious
period, follow specified parametric distributions; and (iii) assuming that any vaccinated individual is actually protected
from disease with a probability that may depend on the characteristics of that individual. If such models are then
fitted to data, we can obtain estimates of quantities of interest such as infection rates, length of infectious
period, measures of vaccine efficacy, epidemiological quantities such as the basic reproduction number, mutation rates
of viruses, and so on. Such estimates are often be used for predictive purposes for potential future outbreaks, for
example for estimating the likely size of such an outbreak, estimating the fraction of a population that should be vaccinated
to prevent such outbreaks, designing optimal control or mitigation strategies, estimating the
healthcare resources required for future epidemics, or evaluating the likely efficacy of proposed public health interventions such
as travel restrictions or school closures.

Given an epidemic model, with parametric models underlying its definition, the typical inference problem
involves estimating the parameters associated with the parametric models from the data to hand. This in itself
is often a non-trivial exercise, due in large part to the fact that in the vast majority of real-life infectious
diseases we do not actually observe the process of infection. Thus on the one hand we have a model whose primary
function is to describe the infection process, but on the other hand the infection process is seldom observed.
Methods to overcome this issue are described elsewhere in this issue, notable examples including the use of
computational approaches such as data-augmented Markov chain Monte Carlo (MCMC) methods \citep{oneill-roberts}
or Approximate Bayesian Computation \citep{cook09}.

From a statistical perspective, it is natural to ask how far the underlying modelling assumptions influence the results
of any analysis. In practice, this question might be addressed by considering alternative parametric models, or
sensitivity analyses if some of the underlying model parameters are assumed to be known {\em a priori}. An alternative
approach is to define the model, or some parts of the model, in a {\em nonparametric} manner. There are two key reasons to do
this. The first is that such an approach typically makes less rigid assumptions than those commonly assumed in epidemic
modelling. In contrast to some simple physical systems, there are many aspects of real-life epidemiology for
infectious diseases that are not well-understood. For instance, in human diseases the actual process of potential
infectious contact between individuals is often hard to precisely describe, as is the detail of variations between individuals
in terms of their susceptibility or infectivity. It is therefore natural to consider models that try to avoid making
potentially restrictive assumptions. The second motivating reason is that fitting a nonparametric model to data provides,
at least informally, some idea of how appropriate particular parametric assumptions are. For example, fitting a nonparametric model
of an infectious period distribution to data might reveal that a particular simple parametric form would provide an adequate model.
Within the context of epidemic modelling, models themselves are often used to explore what-if scenarios or predict future potential
outbreaks, and in both settings it is useful to be able to do so using the simplest possible parametric model. Nonparametric
modelling and inference provides some way of deciding what such a model should be.

\subsection{Classical approaches}
Nonparametric approaches to fitting epidemic models to data have received relatively little attention in the literature.
We now briefly mention some key papers in non-Bayesian settings. Both \citet{becker-yip} and \citet{becker89} describe methods for nonparametric estimation of
the infection rate in the so-called general epidemic model (i.e. the Susceptible-Infective-Removed model with infectious periods distributed
according to an exponential distribution) by allowing the infection rate to depend on time. In doing so they require
that the infection times are known, which as mentioned above is rarely the case in reality.
Conversely, \citet{lau-yip} assume only removal times are observed, and use a kernel estimator to estimate the unobserved
process of infectives, assuming that the parameter of the exponential infectious period distribution
is known. Finally \citet{chen} use kernel estimation to estimate the infection rate in a large-scale epidemic model
in which the depletion of susceptibles is ignored.

\subsection{Layout of this article}
In this article we will focus exclusively on Bayesian nonparametric methods for estimating aspects of the
infection process. In section \ref{sec:prelim} we recall relevant background material. Section \ref{sec:cts}
reviews some Bayesian nonparametric continuous-time stochastic epidemic models and associated methods of inference.
In section \ref{sec:disc} we then develop corresponding models and methods for discrete-time models. In the context
of infectious disease data analysis, discrete-time models are often very natural since real-life data are invariably
discrete in time (e.g. the number of observed cases each day or week). These methods have not appeared in the
literature before and so we illustrate them with various examples. We finish with some concluding remarks in section \ref{sec:conclusions}.

\section{Preliminary material}
\label{sec:prelim}
\subsection{The SIR model in continuous time}
In this article we will focus exclusively on models of the Susceptible-Infective-Removed (SIR) type, although the
methods can be applied to more complex epidemic models. The single-population continuous-time SIR model is defined as follows
\citep[see e.g.][]{bailey, andbritt2000}.

Consider a population consisting of $N$ individuals. At any time $t \geq 0$, each member of the population is either
{\em susceptible}, meaning they are capable of contracting the disease in question, {\em infective}, meaning that they have
the disease and can pass it on to others, or {\em removed}, meaning that they are no longer able to infect others and no
longer able to be re-infected. The precise interpretation of the removed state depends on the disease under consideration,
examples including isolation, recovery, or death. At time $t=0$ the population is entirely susceptible apart from a few
infective individuals. Each infective individual remains so for a period of time, known as the infectious period, that has
a pre-specified distribution. In this article we will only consider the case where the infectious period is
exponentially distributed with mean $\gamma^{-1}$. At the end of its infectious period, the individual becomes removed.
The infectious periods of different individuals are assumed to be independent. During its infectious period, a given infective
individual has contacts with any given susceptible in the population at times given by the points of a Poisson process of
rate $\beta$. All such Poisson processes are independent. Any contact that occurs results in the susceptible
individual immediately becoming infective. The epidemic continues until there are no infectives remaining. Thus at the end of the
epidemic, all individuals are either susceptible by virtue of having avoided infection, or removed. Finally, the population is
assumed to be closed in the sense that no individuals may enter or leave during the epidemic. Such an assumption is reasonable
for real-life outbreaks where the epidemic dynamics are much faster than demographic changes such as births, deaths from causes
unrelated to the epidemic, or movements of individuals in and out of the population.

For $t \geq 0$, let $X(t)$ and $Y(t)$ denote, respectively, the numbers of susceptible and infective individuals in the
population at time $t$. The facts that (i) infections occur according to independent Poisson processes and (ii) infectious
periods are independent exponential distributions together imply that the process $\left\{ (X(t), Y(t)): t \geq 0 \right\}$
is a bivariate continuous-time Markov chain \citep{andbritt2000}. Specifically, the transitions are
\begin{eqnarray*}
P[ (X(t+\delta t), Y(t+\delta t)) = (x-1,y+1) | (X(t),Y(t)) = (x,y)] & = & \beta xy \delta t + o(\delta t),\\
P[ (X(t+\delta t), Y(t+\delta t)) = (x,y-1) | (X(t),Y(t)) = (x,y)] & = & \gamma y \delta t + o(\delta t),
\end{eqnarray*}
which correspond respectively to an infection and a removal. The parameters $\beta$ and $\gamma$ are known as the
{\em infection rate} and {\em removal rate}, respectively. We will also refer to $\beta X(t) Y(t)$ as the {\em incidence rate}.

Finally, although the SIR model with exponential infectious periods can be viewed as a Markov chain, this fact in itself is not
important for the Bayesian nonparametric methods that we describe below. Our focus on the particular model defined above is only
for ease of exposition; more general non-Markov epidemic models can also be analysed using the methods we describe.

\subsection{The SIR model in discrete time}
A discrete-time version of the SIR model can be defined similarly to the continuous-time version. At any time
$t=0, 1, 2, \ldots$, the population is divided into susceptibles, infectives and removed individuals. The infectious
period distribution is a positive integer-valued random variable. An individual who becomes infective at time $t$ and
whose infectious period is of $k$ time units becomes removed at time $t+k$. At any time $t$, each susceptible individual
independently avoids infection with probability $\exp \left\{- \beta Y(t) \right\}$. Those failing to avoid infection become infective
at time $t+1$. The epidemic continues until no infectives remain.

Note that the infection process in this model arises by approximating the Poisson process assumption in the continuous-time
model. Specifically, in the latter the probability of a given susceptible avoiding infection for one time unit is
$\exp \left\{- \beta Y(t) \right\}$, assuming that no other events occur. Thus the discrete-time model provides a good approximation
to the continuous-time model if the time units are sufficiently small.

\subsection{Time-dependent infection process}
A natural generalisation of the SIR model is to suppose that the infection rate $\beta$ is time-dependent. This
could be done to describe genuine changes in population mixing over time, or as a proxy for some unobserved
heterogeneity in the population that gives rise to an apparent change in infection rate over time
\citep[see e.g.][]{Lekone,pollicott12,smirnova-tuncer}.

In terms of inference, we may therefore wish to estimate $\beta(t)$ for all time points $t$. If we assume a parametric
model for $\beta(t)$ then this typically introduces a small number of extra model parameters. However, if we
adopt a nonparametric approach in which we do not impose a particular parametric structure, then estimating
$\beta(t)$ for all $t$ amounts to estimating an uncountably-infinite-dimensional object in the continuous-time
case; and a finite- or countably-infinite-dimensional object in the discrete-time case, depending on whether
or not the infectious period distribution has finite support.

An alternative kind of generalisation is to allow the incidence rate to be time-dependent, so that we replace
$\beta X(t) Y(t)$ with $\beta(t)$, or more realistically $\beta(t) \chi_{ \left\{ X(t)Y(t) \geq 1 \right\} }$,
where $\chi_A$ is the indicator function of the event $A$. The latter formulation ensures that new infections
can only occur if there is at least one infective and one susceptible in the population. The motivation for
such a model is to relax the usual assumption that infections occur at a rate proportional to $X(t)Y(t)$, which
itself essentially arises by assuming that the individuals in the population mix together homogeneously.
Another motivation for this kind of model is that it can act as a baseline
case in any analysis that involves comparing different models.

\subsection{Bayesian inference for the standard continuous-time SIR model}
The nonparametric methods described in the next section involve appropriate modifications of a common
approach to inference for the parametric SIR model. For this reason, we now recall the latter, as described in \cite{oneill-roberts}.

The basic problem is to infer the infection and removal rates in an SIR model, given that only removals are
observed. However, even for the SIR model with fixed infection rate the likelihood of observed removals
given the model parameters is intractable in all but the simplest cases. This is because calculating the
likelihood involves integrating over the space of all possible infection times. Although in principle this is
possible, in practice it is a very cumbersome and computationally expensive approach. An attractive alternative
is to use data augmentation in an MCMC setting, specifically including the unobserved infection times as extra
variables.

Consider a continuous-time SIR model with infection rate $\beta$, removal rate $\gamma$,
a population of $N$ individuals, and one initial infective. Suppose we observe removals at times
$r_1, \ldots, r_n$, where $r_1 < r_2 < \ldots < r_n < T$, so that we observe the epidemic until time $T$.
Denote by $i_1, \ldots, i_m$ the ordered unobserved infection times in $(-\infty, T]$ , where $m \geq n$, so that
$i_1 < i_2 < \ldots < i_m$. Define $\mathbf{r} = (r_1, \ldots, r_n)$ and
$\mathbf{i} = (i_2, \ldots, i_m)$. Then the augmented likelihood of infection and removal times is
\begin{eqnarray}
\label{SIR_auglike}
\lefteqn{\mbox{} \hspace{0.5cm}\pi ( \mathbf{i}, \mathbf{r} | i_1, \beta, \gamma )}  \\
&& = \prod_{j=2}^m \beta X(i_j - ) Y(i_j - )
\prod_{j=1}^n \gamma Y(r_j - ) \exp \left\{ - \int_{i_1}^{T} \beta X(t)Y(t) + \gamma Y(t) \; dt  \right\}, \nonumber
\end{eqnarray}
where $X(i_j -) = \lim_{s \uparrow i_j} X(s)$, etc.

A brief explanation for (\ref{SIR_auglike}) is as follows (for more
detailed accounts see \cite{andbritt2000} for SIR models, or \cite{andersen1993} for more general counting process
models, of which the SIR model is a special case). First note that in a small time interval $[t, t + \delta t)$,
the probabilities of an infection, a removal, and no event are approximately $\beta X(t) Y(t) \delta t$, $\gamma Y(t) \delta t$
and $1 - (\beta X(t) Y(t) + \gamma Y(t)) \delta t$, respectively. By splitting the time interval $(i_1, T]$
into a large number of such small intervals, the probability of the observed process is thus given by a product of
such probabilities, all but $m+n-1$ of which correspond to no event occurring. Note here that we do not include the
infection at time $i_1$, since this is an assumed initial condition, so there is no contribution to the likelihood.
As $\delta t \downarrow 0$, and moving from probability to density, the terms that remain are (i) two product terms corresponding to the infection and
removal events, and (ii) an exponential term which arises since $1 - (\beta X(t) Y(t) + \gamma Y(t)) \delta t \approx
\exp\left\{ - (\beta X(t) Y(t) + \gamma Y(t)) \delta t \right\}$.

The object of interest is the augmented joint posterior density of $\beta$, $\gamma$ and the unobserved infection
times. From Bayes' Theorem we have
\begin{equation}
\label{SIR_post}
\pi ( \mathbf{i}, i_1, \beta, \gamma | \mathbf{r}) \propto \pi ( \mathbf{i}, \mathbf{r} | i_1, \beta, \gamma ) \pi (\beta, \gamma, i_1).
\end{equation}
Samples from the target density can be obtained via MCMC, as follows. In practice we usually assign independent prior distributions to
$\beta$, $\gamma$ and $i_1$. If $\beta$ and $\gamma$ are assigned gamma prior distributions, it follows from (\ref{SIR_auglike}) and (\ref{SIR_post}) that
both have gamma-distributed full conditional distributions. This in turn means that both parameters can be updated using Gibbs
steps within an MCMC algorithm. To assign a prior distribution for $i_1$, observe that $i_1 < r_1$ and assign a prior distribution
to $r_1 - i_1$. If this distribution is exponential then the full conditional distribution for $i_1$ is tractable, which again means
that $i_1$ can be updated using a Gibbs step.

Finally, the infection times $i_2, \ldots, i_m$ can be updated using Metropolis-Hastings steps.
These consist of adding, deleting and moving infection times, although if the epidemic is known to have ceased by time $T$ then $m=n$
and only the third of these updates is needed. In practice, it is useful to perform a number of such updates during each iteration
of the MCMC algorithm, in order to improve the mixing of the Markov chain.
Each individual update involves proposing to add, delete or move an infection time, where new infections or moves might involve
proposing times uniformly on the range of possible values, and then evaluating the usual Metropolis-Hastings acceptance probability
using (\ref{SIR_post}). Full details of such an algorithm, which technically speaking is a reversible-jump MCMC algorithm, are
given in \cite{oneill-roberts}.

\section{Nonparametrics for continuous-time SIR models}
\label{sec:cts}
In this section we describe several approaches to Bayesian nonparametric modelling for parameters
associated with the infection process in an SIR model. Specifically, we consider models in which
either the infection rate $\beta$ or the incidence rate $\beta X(t) Y(t)$ is replaced by a
time-dependent quantity. This quantity can be assigned a prior model in various different ways as
described below, and this assignment in turn determines the kind of MCMC algorithm that is required for inference.
In each case it is assumed that the data to hand consist of removal times, and that infectious periods
are exponentially distributed. For ease of exposition, we also assume that the epidemic has been completed
during the observation period, so that the number of unobserved infections ($m$) equals the number of observed removals ($n)$.
We may also set $T=r_n$ since no further events occur after time $r_n$.

To begin with, suppose we replace the infection rate $\beta$ in the SIR model with the time-dependent version
$\beta(t)$. The modified likelihood corresponding to (\ref{SIR_auglike}) is
\begin{eqnarray}
\label{SIR_infrate_like}
\pi ( \mathbf{i}, \mathbf{r} | i_1, \beta, \gamma ) & = & \prod_{j=2}^n \beta (i_j - ) X(i_j - ) Y(i_j - )
\prod_{j=1}^n \gamma Y(r_j - )  \\
&& \times \exp \left\{ - \int_{i_1}^{r_n} \beta(t) X(t)Y(t) + \gamma Y(t) \; dt  \right\}, \nonumber
\end{eqnarray}
where now $\beta$ denotes the function whose value at $t$ is $\beta(t)$. The difficulty that now arises is
that the integral term in (\ref{SIR_infrate_like}) involves the infinite-dimensional object $\beta$,
and is hence intractable if we do not assume a particular parametric form for $\beta$.

Methods for overcoming this problem depend on the prior structure we impose on $\beta$. For instance, if we
model $\beta$ using a step function, it can then be defined using only finitely many values, which in turn yields
a tractable MCMC scheme to infer $\beta$. Conversely if we impose a Gaussian Process prior structure then there is no such
dimensionality reduction, and we need an alternative method, as we now describe.

\subsection{Background on Gaussian process methods}
We first present relevant facts concerning Gaussian Processes (GPs). A comprehensive account can be found in \cite{rasmussen06}.
Recall that a Gaussian Process (GP) is a stochastic process whose realisations consist of Gaussian random variables indexed
by some set. In our case, the latter will be the set of times $t$ in some interval. A GP is completely specified by its mean
and covariance function. In Bayesian nonparametrics, GPs are commonly used as prior models for functions. For
example, assigning a GP prior to a function $f:[0, \infty ] \rightarrow \mathbb{R}$ means that for any $x_1, \ldots, x_n \geq 0$,
the vector $(f(x_1), \ldots, f(x_n))$ has a multivariate Gaussian distribution with mean vector $\mu(x_1, \ldots, x_n)$ and covariance matrix
$\Sigma(x_1, \ldots, x_n)$. In many situations, it is common to use zero-mean GPs, since offsets can often easily be removed before
modelling starts.

In our setting, we wish to assign a prior to the time-dependent infection rate $\beta$. Since $\beta(t) \geq 0$, we cannot do this
directly since a GP model would assign positive probability to the event $\beta(t) < 0$. We therefore use a transformation, the
details of which are given below.

As described in \cite{xuetal16}, one way to deal with the intractable integral in (\ref{SIR_infrate_like}) is to
exploit the fact that a time-inhomogeneous Poisson process can be constructed using a suitable thinned homogeneous Poisson
process. This approach is used in \cite{adams-murray09} to provide a method for Bayesian nonparametric inference for a
time-inhomogeneous Poisson process. We now recall the details of the latter.

Suppose we observe a set of points $\mathbf{s} = (s_1, \ldots, s_K)$ from a Poisson process with time-dependent intensity $\lambda(t)$ during $[0,T]$. The likelihood of
these observations is
\[
\pi( \mathbf{s} | \lambda) =   \prod_{k=1}^K \lambda(s_k-)  \exp \left\{ - \int_0^T \lambda(t) \; \mathrm{d}t \right\},
\]
and as for the epidemic case the integral is intractable.

The key idea is that the original process can be viewed as a thinned
homogeneous Poisson process of rate $\lambda^*$, where $\lambda(t) \leq \lambda^*$ for all $0 \leq t \leq T$, in which a point at time $t$ is retained with
probability $\lambda(t)/\lambda^*$ . We may thus augment the observed data with the unobserved thinned points, $\tilde{\mathbf{s}}= (\tilde{s}_1, \ldots, \tilde{s}_M)$, say, yielding an augmented likelihood
\[
\pi( \mathbf{s}, M, \tilde{\mathbf{s}} | \lambda, \lambda^*) = (\lambda^{*})^{M+K}\exp\{-\lambda^{*}T\}
 \prod_{k=1}^K \frac{\lambda(s_k-)}{\lambda^*}  \prod_{m=1}^M \left( 1 - \frac{\lambda(\tilde{s}_m-)}{\lambda^*} \right) .
\]
This new likelihood can evidently be computed with only finitely many evaluations of the function $\lambda$.

Next, suppose we wish to impose a GP prior on $\lambda$. Since $\lambda \geq 0$ this cannot be done directly, and so
instead we use the transformation $\lambda(t) = {\lambda}^{*}{\sigma}(g(t))$, where ${\sigma}(z)=(1+e^{-z})^{-1}$, and assign a GP prior distribution to $g$. This prior distribution is specified by assuming a particular form of covariance function with parameter vector $\theta$, and placing a prior distribution on $\theta$.

Let ${\bf g}_{M+K} = (g(s_{1}-),g(s_{2}-),\cdots,g(s_{K}-), g(\tilde{s}_{1}-),g(\tilde{s}_{2}-),\cdots,g(\tilde{s}_{M}-))$. Then from Bayes' Theorem the augmented posterior density of interest is
\begin{eqnarray*}
\pi(g, \lambda^*, M, \tilde{\mathbf{s}}, \theta | \mathbf{s} ) & \propto &
		  (\lambda^{*})^{M+K}\exp\{-\lambda^{*}T\}\prod_{k=1}^{K}\sigma(g(s_{k}-))\prod_{m=1}^{M}\sigma(-g(\tilde{s}_m-))\\
&&\times \, {\pi({\bf g}_{M+K}| \theta)} \pi(\lambda^*) \pi (\theta),
\end{eqnarray*}
where $\pi({\bf g}_{M+K} | \theta)$ is the density of a multivariate Gaussian random variable,
and $\pi(\lambda^*)$ and $\pi(\theta)$ are respectively the prior density functions of $\lambda^*$ and $\theta$, assuming prior
independence. The posterior density can be explored using MCMC methods, and since $\lambda$ is specified by $g$
and $\lambda^*$, we may hence obtain posterior samples for $\lambda$.

\subsection{Gaussian process methods for the SIR model}
\label{subsec:GP for cts time}
\cite{xuetal16} describe how to adapt the ideas above to the SIR model in continuous time, as follows. With notation as before,
we observe removal times $\mathbf{r}$ and wish to infer the infection rate function $\beta$ and the removal rate parameter $\gamma$.
As for the parametric case, we introduce the initial infection time $i_1$ and subsequent infection times $\mathbf{i}$.

The key idea is to observe that the rate of infections at time $t$ is $\beta (t) X(t)Y(t)$, and the infection process can be
constructed by thinning a bounding process of rate $\beta^* X(t)Y(t)$, provided $\beta(t) \leq \beta^*$ for all $t$.
To make this precise, we may clearly construct a realisation of the epidemic by generating a sequence of inter-event times and event types.
Suppose an event has just occurred at time $t$, and $(X(t),Y(t)) = (x,y)$. First simulate the potential time until the next removal event, $\tau_R$ say,
which has an exponential distribution with mean $(\gamma y)^{-1}$.
Next construct the potential time until the next infection event, $\tau_I$, by simulating a Poisson process of
rate $\beta^* xy$, thinning it by independently retaining each point at time $s$ with probability $\beta(s) / \beta^*$, and setting
$\tau_I$ as the time until the first retained point appears. Finally, $\min(\tau_R, \tau_I)$ is the time until the next
event in the epidemic, which is an infection if $\tau_I < \tau_R$ and a removal otherwise. Iterating this procedure provides a
realisation of the epidemic, ending as soon as $Y(t) = 0$.

The above construction enables us to write down the joint likelihood of the bounding process and the resulting
realisation, since the ingredients just consist of homogeneous Poisson processes and Bernoulli trials. Specifically,
consider the likelihood of what occurs during $(t,t+\tau]$, where as above an event has just occurred
at time $t$, and $\tau$ is the time until the next event. Suppose that thinned events occur in $(t,t+\tau)$ at times $s_1, \ldots, s_m$,
where $m=0$ if no thinned events occur. These events give a likelihood contribution
\begin{equation}
\label{thinned}
\prod_{j=1}^m \beta^* X(s_j) Y(s_j) (1 - \beta(s_j)/\beta^*) \exp \left( - \int_t^{t+\tau} \beta^* X(u) Y(u) \; du \right) ,
\end{equation}
where the product term equals 1 if $m=0$. Next, if the event at $t+\tau$ is an infection then this generates (i) the additional product term
$\beta^* X(t+\tau) Y(t+\tau) (\beta(t+\tau)/\beta^*)$ which corresponds to a non-thinned event and (ii)
the probability that $\tau_R > \tau$, namely $\exp( - \int_t^{t+\tau} \gamma Y(u) \; du)$. Note that here we are essentially
integrating out the value of $\tau_R$ from the construction, i.e. the observed event at $t+\tau$ could have
arisen for any value of $\tau_R$ greater than $\tau$. Finally, if the event at $t+\tau$ is a removal then
this generates the likelihood contribution that $\tau_R = \tau$, namely $\gamma Y(t+\tau)  \exp( - \int_t^{t+\tau} \gamma Y(u) \; du)$.
Note that we do not require any further term for the infection process, since the exponential term in (\ref{thinned}) essentially
integrates out all possible $\tau_I > \tau$.

Following the methods of inference for inhomogenous Poisson processes described above, we then assign a prior model for $\beta$ by
setting $\beta(t) = \beta^* \sigma(g(t))$. Here, $g$ is a random function drawn from a GP with a specified covariance
function with parameter $\theta$.
We require additional variables, namely the number of thinned events, $M$; their locations, $\tilde{\mathbf{i}} = (\tilde{i}_1, \ldots, \tilde{i}_M)$; the $g$ function values at the infection times, $\mathbf{g}_n = (g(i_2-), \ldots, g(i_n-))$, and the $g$ function values at the thinned event times, $\mathbf{g}_M = (g(\tilde{i}_1-), \ldots, g(\tilde{i}_M-))$.
The augmented likelihood is
\begin{eqnarray*}
\lefteqn{\pi( \mathbf{i}, \mathbf{r}, M, \tilde{\mathbf{i}} | \beta^*, \theta, \gamma, i_1, \mathbf{g}_n, \mathbf{g}_M )} \\
&=& \prod_{j=2}^{n} \beta^* X(i_j-)Y(i_j-) \sigma(g(i_j-))
\prod_{j=1}^{M} \beta^* X(\tilde{i}_j-)Y(\tilde{i}_j-) \sigma(-g(\tilde{i}_j-))
\prod_{j=1}^{n} \gamma Y(r_j-)\\
&& \times \exp \left( -\int_{i_1}^{r_n} \beta^* X(t) Y(t) + \gamma Y(t) \; dt \right),
\end{eqnarray*}
and the posterior target density is
\begin{eqnarray}
\label{SIR_infrate_posterior}
\lefteqn{\pi ( g, \beta^*, \gamma, \mathbf{i}, i_1, M, \tilde{\mathbf{i}}, \theta | \mathbf{r} ) \propto }\\
&& \pi( \mathbf{i}, \mathbf{r}, M, \tilde{\mathbf{i}} | \beta^*, \theta, \gamma, i_1, \mathbf{g}_n, \mathbf{g}_M )
\pi ( \mathbf{g}_n, \mathbf{g}_M | \theta) \pi(\theta) \pi (\beta^*) \pi(\gamma) \pi (i_1), \nonumber
\end{eqnarray}
assuming prior independence for the model parameters. Samples from (\ref{SIR_infrate_posterior}) can be obtained using an MCMC algorithm
similar to that described above for the parametric SIR case, but now incorporating the additional parameters associated
with the thinned points and the GP. Full details can be found in \cite{xuetal16}.

As mentioned above, instead of modelling the infection rate in a nonparametric manner, one could also model the incidence
rate. In this case, $\beta X(t) Y(t)$ is replaced by a single function $\beta(t) \chi_{ \left\{ X(t)Y(t) \geq 1 \right\} }$.
The GP approach described above can easily be adapted to this setting.

\subsection{Step-function and B-spline methods}
Another way to assign a prior to the infection or incidence rate functions is to use step functions. \cite{knock} do this
by assuming that the incidence rate function is of the form $\beta(t) \chi_{ \left\{ X(t)Y(t) \geq 1 \right\} }$, where
$\beta$ is a step function with change-points $s_1 < s_2 < \ldots < s_k$, so that
\[
\beta(t) = \sum_{j=0}^k \beta_j \chi_{ \left\{ s_j \leq t < s_{j+1} \right\} }
\]
where $s_0 = i_1$ and $s_{k+1} = r_n$. Under this assumption, the integral in (\ref{SIR_auglike}) can be easily evaluated
and so the augmented likelihood is tractable. Note that the number of change-points is itself not assumed to be known.
The inference problem for $\beta$ thus reduces to estimation of the number and location of the change-points, and the function values
$\beta_0, \ldots, \beta_k$. The posterior target density is
\[
\pi ( \beta, k, s, \gamma, \mathbf{i}, i_1 | \mathbf{r} )  \propto \pi ( \mathbf{i}, \mathbf{r} | \beta, k, s, \gamma, i_1)
\pi (\beta, k, s, \gamma, i_1),
\]
where $\beta$ and $s$ denote respectively the function values and change-point locations. Options for assigning prior distributions
are described in \cite{knock}; for the function values these include independent priors for each $\beta_j$, and also sequentially
dependent priors in which $E[ \beta_{j+1} | \beta_j ] = \beta_j$. The latter essentially gives some element of smoothing to the
incidence rate function and is thus similar in spirit to the GP methods described above.

A related approach described in \cite{knock} is to assume that the incidence rate is piece-wise quadratic, specifically that
it can be modelled as a second-order B-spline. As for the step function case, this yields a tractable augmented likelihood,
and inference involves estimating the parameters of the B-spline function.

\section{Nonparametrics for discrete-time SIR models}
In this section we show how to adapt the Gaussian process methods described above to the scenario of a discrete-time epidemic
model. We illustrate the methods with some examples.
\subsection{Methods}
\label{sec:disc}
Consider a discrete-time SIR model in which the infection rate at integer time $t$ is
denoted $\beta(t)$, and in which infectious periods have a distribution with
probability mass function $p(k | \eta)$, $k = 1, 2, \ldots$, where $\eta$ is the parameter vector of
the distribution.  Suppose we observe removals at integer times
$\mathbf{r} = (r_1, \ldots, r_n)$, where $r_1 \leq r_2 \leq \ldots \leq r_n$ and we label the
corresponding individuals $1, \ldots, n$. Let $i_j$ denote the time that individual $j$ starts being infective.
This means that if individual $j$ is susceptible at time $t$ and fails to avoid infection, they become
infective at time $t+1$ and so $i_j = t+1$. Since they are removed at time $r_j$, their infectious
period is $r_j - i_j$. Let $\kappa$ denote the label of the initial infective individual,
i.e. $i_\kappa < i_j$ for all $j \neq \kappa$. We only allow one initial infective, although this assumption
could be relaxed. For simplicity in exposition we also assume that the epidemic is known to have finished, so
that $i_1, \ldots, i_n$ are the only infection times.

The augmented likelihood corresponding to (\ref{SIR_auglike}) is
\begin{eqnarray}
\label{SIR_disc_auglike}
\lefteqn{\pi ( \mathbf{i}, \mathbf{r} | i_\kappa, \kappa, \beta, \eta )
= \prod_{j \neq \kappa} \left( 1 - \exp \left\{ - \beta (i_j - 1) Y(i_j -1 ) \right\} \right) }\\
&& \times \exp \left\{ - \sum_{t = i_\kappa}^{r_n - 1} \beta(t) Y(t) X(t+1) \right\}
\prod_{j=1}^n p(r_j - i_j | \eta). \nonumber
\end{eqnarray}
The methods described above using Gaussian processes can be readily extended to this situation, as follows.
First, we assume {\em a priori} that $\beta = \exp(g(t))$, where $g$ is drawn from a zero-mean GP prior
whose covariance function has parameter vector $\theta$. Set $\mathbf{g} = (g(i_\kappa), \ldots, g(r_n - 1))$.
The posterior density of interest is then
\[
\pi (g, \eta, \mathbf{i}, i_\kappa, \kappa | \mathbf{r}) \propto \pi ( \mathbf{i}, \mathbf{r} | i_\kappa, \kappa, \mathbf{g}, \eta )
\pi (\mathbf{g} | \theta) \pi (\theta) \pi (\eta) \pi (i_\kappa) \pi(\kappa),
\]
where the likelihood $\pi ( \mathbf{i}, \mathbf{r} | i_\kappa, \kappa, \mathbf{g}, \eta )$ is obtained by replacing
$\beta(t)$ with $\exp(g(t))$ in (\ref{SIR_disc_auglike}). This target density can then be explored using MCMC methods.

\subsection{Examples}
We now illustrate the methods using both simulated data and real data. We assume that the infectious period
distribution is geometric with mean $\gamma^{-1}$, this being the discrete analogue of the exponential infectious periods
adopted for the continuous-time models in Section 3. Let $\gamma \sim Beta(\lambda_\gamma, \nu_\gamma)$ {\em a priori},
and for simplicity we fix the GP covariance function parameter $\theta$. We assign a uniform prior distribution on $\left\{ 1, \ldots, n \right\}$ to $\kappa$, and an improper uniform distribution on $\left\{ t : t = r_1-1, r_1-2, \ldots \right\}$ to $i_\kappa$.
The target density of interest is thus

\begin{eqnarray}
\label{SIR_disc_auglike_example}
\lefteqn{\; \; \; \; \; \; \pi (g, \gamma, \mathbf{i}, i_\kappa, \kappa | \mathbf{r}) \propto
\prod_{j \neq \kappa} \left( 1 - \exp \left\{ - \beta (i_j - 1) Y(i_j -1 ) \right\} \right) }\\
&& \times \exp \left\{ - \sum_{t = i_\kappa}^{r_n - 1} \beta(t) Y(t) X(t+1) \right\}
\gamma^{n + \lambda_\gamma - 1} (1 - \gamma)^{\sum_{j=1}^n (r_j - i_j) - n + \nu_\gamma - 1} \pi (\mathbf{g} | \theta). \nonumber
\end{eqnarray}

{\em MCMC algorithm}
An MCMC algorithm which produces samples from the target density defined at (\ref{SIR_disc_auglike_example}) can be obtained by specifying update mechanisms for each of the parameters. The algorithm itself then proceeds by sampling the parameters in turn.
In practice it is usually beneficial to perform several updates for the unknown infection times during each iteration of the
algorithm, in order to improve the mixing of the Markov chain. We now describe the parameter updates.

First note that, from (\ref{SIR_disc_auglike_example}), the full conditional distribution for $\gamma$ is
\[
\gamma |  g, \mathbf{i}, i_\kappa, \kappa, \mathbf{r} \sim Beta \left( n + \lambda_\gamma, \sum_{j=1}^n (r_j - i_j) - n + \nu_\gamma \right)
\]
and so $\gamma$ can be updated by drawing from this distribution.

We update infection times as follows. First, an individual $j$ is chosen uniformly at random from the $n$ infectives. Next, we propose
a new infection time $\tilde{i}_j = r_j - W$, where $W$ has a geometric distribution with mean $\gamma^{-1}$. Note that if $\tilde{i}_j < i_{\kappa}$
then $j$ is also proposed as the new initial infective, i.e. $\tilde{\kappa} = j$; otherwise, $\tilde{\kappa} = \kappa$. If $j \neq \kappa$ and
$\tilde{i}_j = i_{\kappa}$ then the move is immediately rejected, since it has zero likelihood under the assumption that there is exactly one initial infective. The move is accepted with probability $\min\left\{ 1, h(\tilde{\mathbf{i}}, \tilde{i_\kappa}, \tilde{\kappa};g) / h(\mathbf{i}, i_\kappa, \kappa;g) \right\}$, where
$\tilde{\mathbf{i}}$ denotes the proposed set of infection times (i.e. with $i_j$ replaced by $\tilde{i_j}$), and
\begin{eqnarray*}
\lefteqn{h(\mathbf{i}, i_\kappa, \kappa; g) =}\\
&& \prod_{j \neq \kappa} \left( 1 - \exp \left\{ - \beta (i_j - 1) Y(i_j -1 ) \right\} \right)
\times \exp \left\{ - \sum_{t = i_\kappa}^{r_n - 1} \beta(t) Y(t) X(t+1) \right\} .
\end{eqnarray*}

Finally, $g$ is updated by proposing a new value
\[
\tilde{g} = ( \sqrt{1 - \epsilon^2}) g + \epsilon V,
\]
where $V$ denotes an $n$-dimensional Gaussian random variable with mean zero and covariance matrix $\Sigma(\theta)$, and $ 0< \epsilon < 1$ is a tuning
parameter. This so-called under-relaxed proposal method is described in \cite{adams-murray06}. The proposed new value is accepted with probability
$\min \left\{ 1, h(\mathbf{i}, i_\kappa, \kappa; \tilde{g}) / h(\mathbf{i}, i_\kappa, \kappa;g) \right\}$.\\

{\em Simulated data} We used the MCMC algorithm described above to infer the infection rate function in two scenarios: one where
the infection rate decreases slowly over time (Scenario 1), and one where it is periodic (Scenario 2). One data set, consisting of a set of removal times,
was generated for each scenario by simulation from the true model. The data sets shown in the results below were both typical outbreaks.
In both scenarios we set the covariance matrix of the GP to be $\Sigma = (K(x_i,x_j))$ where $K$ is the squared-exponential function
\[
K(x_i,x_j) = \omega \exp \left\{ -\frac{1}{2} \left( \frac{x_i-x_j}{l} \right)^2 \right\}
\]
where $\omega$ and $l$ were chosen to provide reasonably vague prior information for $g$ in each setting. Note that $l$, usually called
the length scale, controls the extent to which the GP can vary over time. Roughly speaking, small values of $l$ allow the GP to vary rapidly, while
larger values only allow slower variation. In the context of epidemic modelling, we might set $l$ to be the time period over which we might
reasonably expect to see little variation in the infection rate. Conversely, $\omega$ controls the variance of the GP at a given input point, akin to
variance of a Gaussian prior distribution on a univariate parameter. Thus larger values correspond to vague prior assumptions in this respect.

We first analysed each simulated data set by assuming that the infection times
were also known. Although this assumption is not very realistic in practice, we do so here to illustrate some features
of the inference problem. With fixed infection times we can easily obtain a maximum likelihood estimate of
$\beta(t)$ on each day of the outbreak, since the known number of new infections each day follows a Binomial distribution.
These estimates can be plotted against the true $\beta(t)$, and this gives some indication of how feasible it is
to estimate the infection rate function. Estimating $\beta$ using our nonparametric approach here (i.e. the
MCMC algorithm, but with infection times fixed at the known values) then illustrates that our GP prior introduces
an element of smoothing compared to the ML estimation. Finally, we then analysed the data without assuming known infection times.

\begin{figure}[ht]
\centering
\subfloat[Removals each day]{%
\includegraphics[height=7cm,width=7cm]{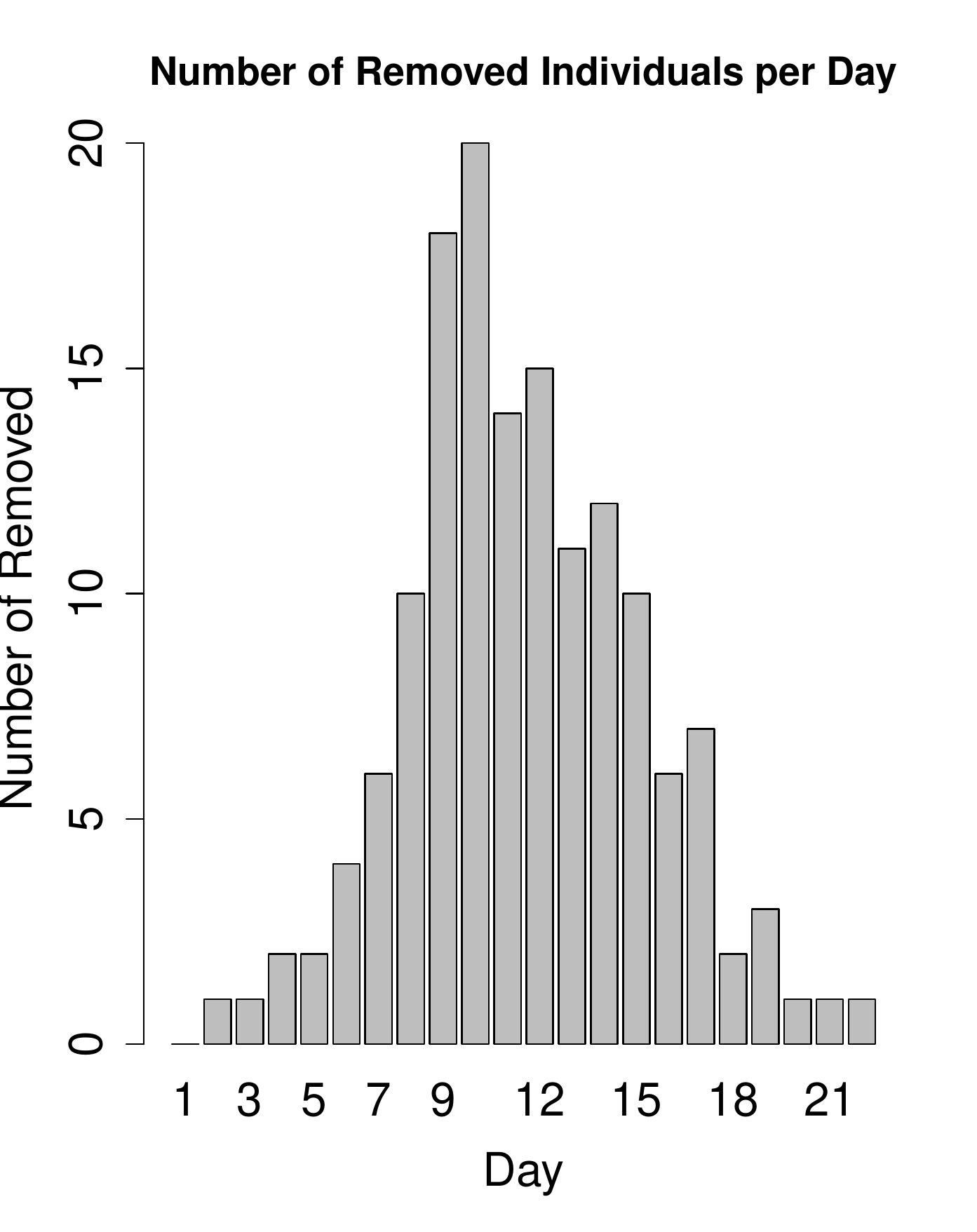}
\label{fig:esubfigure1}}
\subfloat[True $\beta$ and ML estimates]{%
\includegraphics[height=7cm,width=7cm]{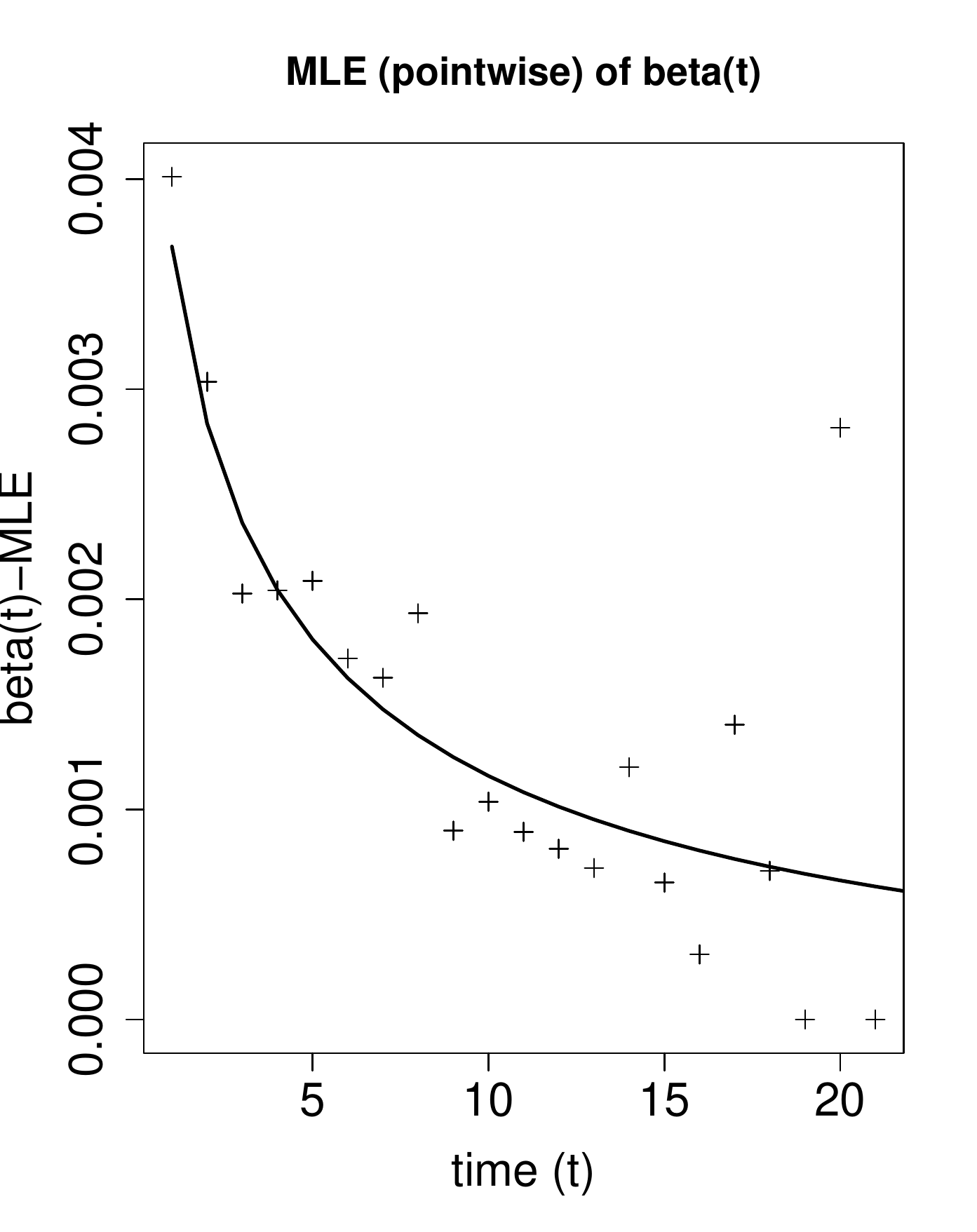}
\label{fig:esubfigure2}}
\quad
\subfloat[Posterior $\beta$, infection times known]{%
\includegraphics[height=7cm,width=7cm]{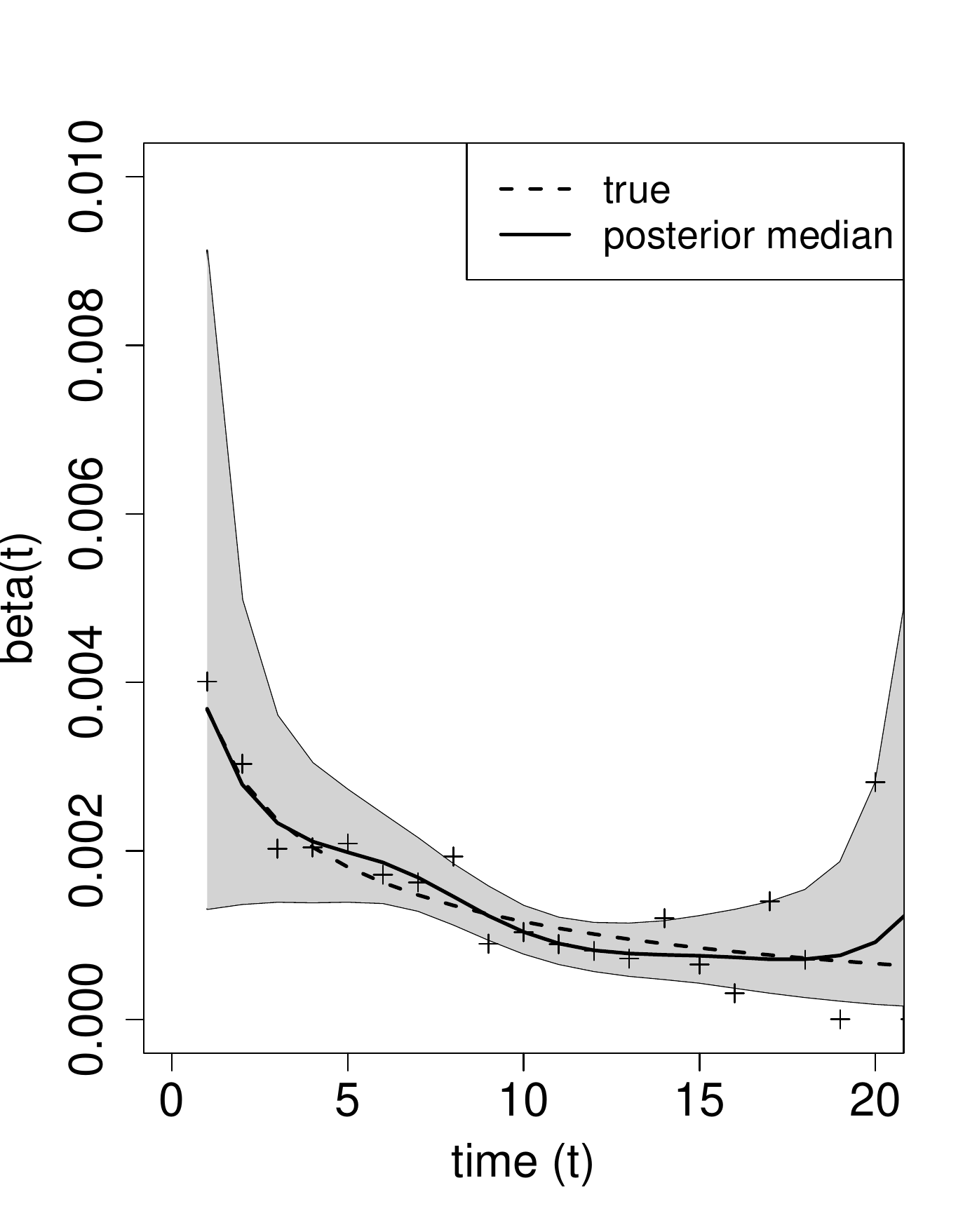}
\label{fig:esubfigure3}}
\subfloat[Posterior $\beta$, infection times unknown]{%
\includegraphics[height=7cm,width=7cm]{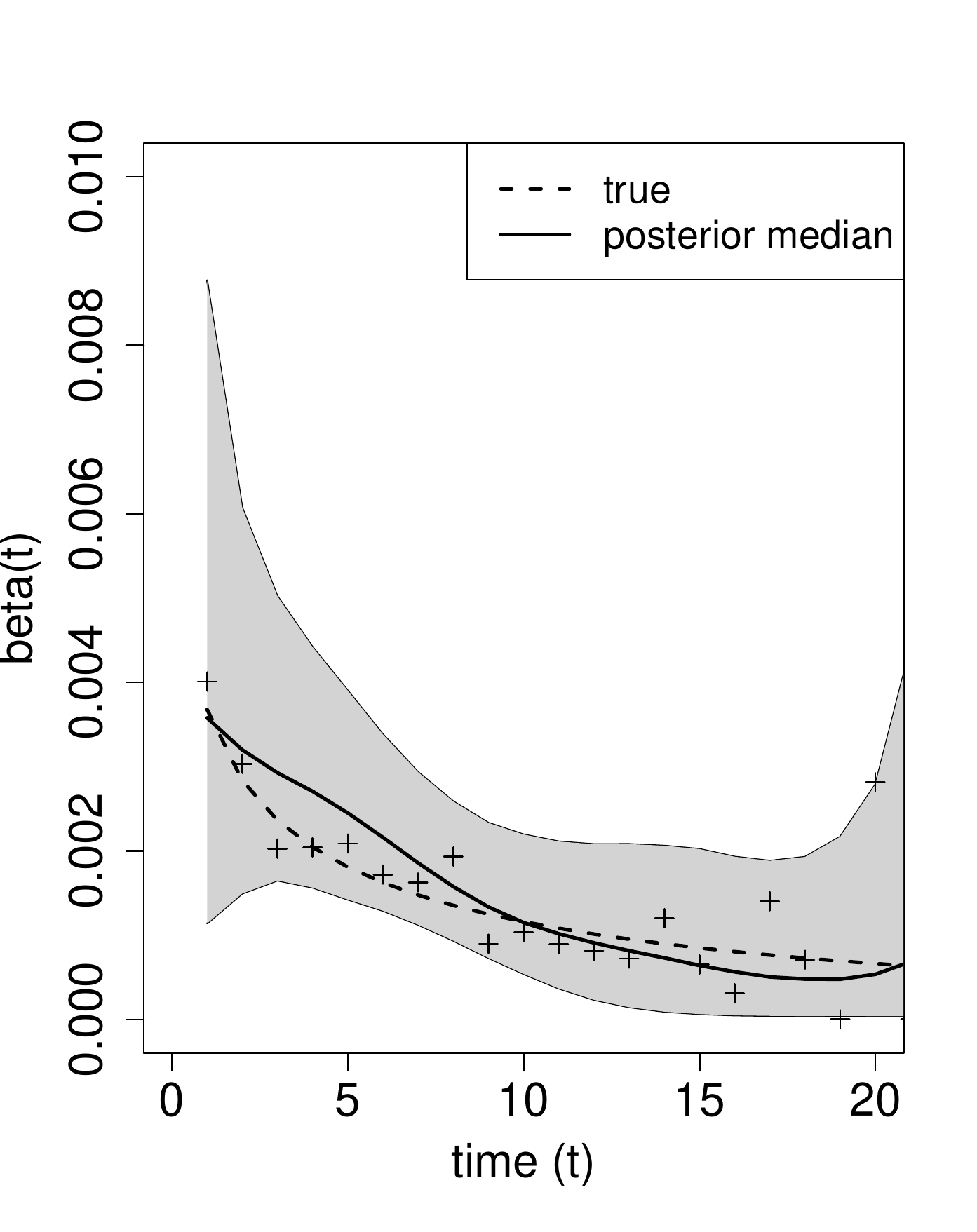}
\label{fig:esubfigure4}}
\caption{Results for simulated data in Scenario 1 using $\beta(t) = (0.01) \exp(-t^{1/3})$, $\gamma = 0.5$, $N=500$, and GP prior covariance
function values $\omega = 10$, $l=6$. Plots (b) and (c) are
obtained using the known true infection times, and plot (d) only uses the removal data. The shaded regions in
(c) and (d) are 95\% posterior credible intervals for $\beta$. The ML estimates are shown as crosses in (c) and (d) for comparison.}
\label{exponential_example}
\end{figure}

\begin{figure}[ht]
\centering
\subfloat[Removals each day]{%
\includegraphics[height=7cm,width=7cm]{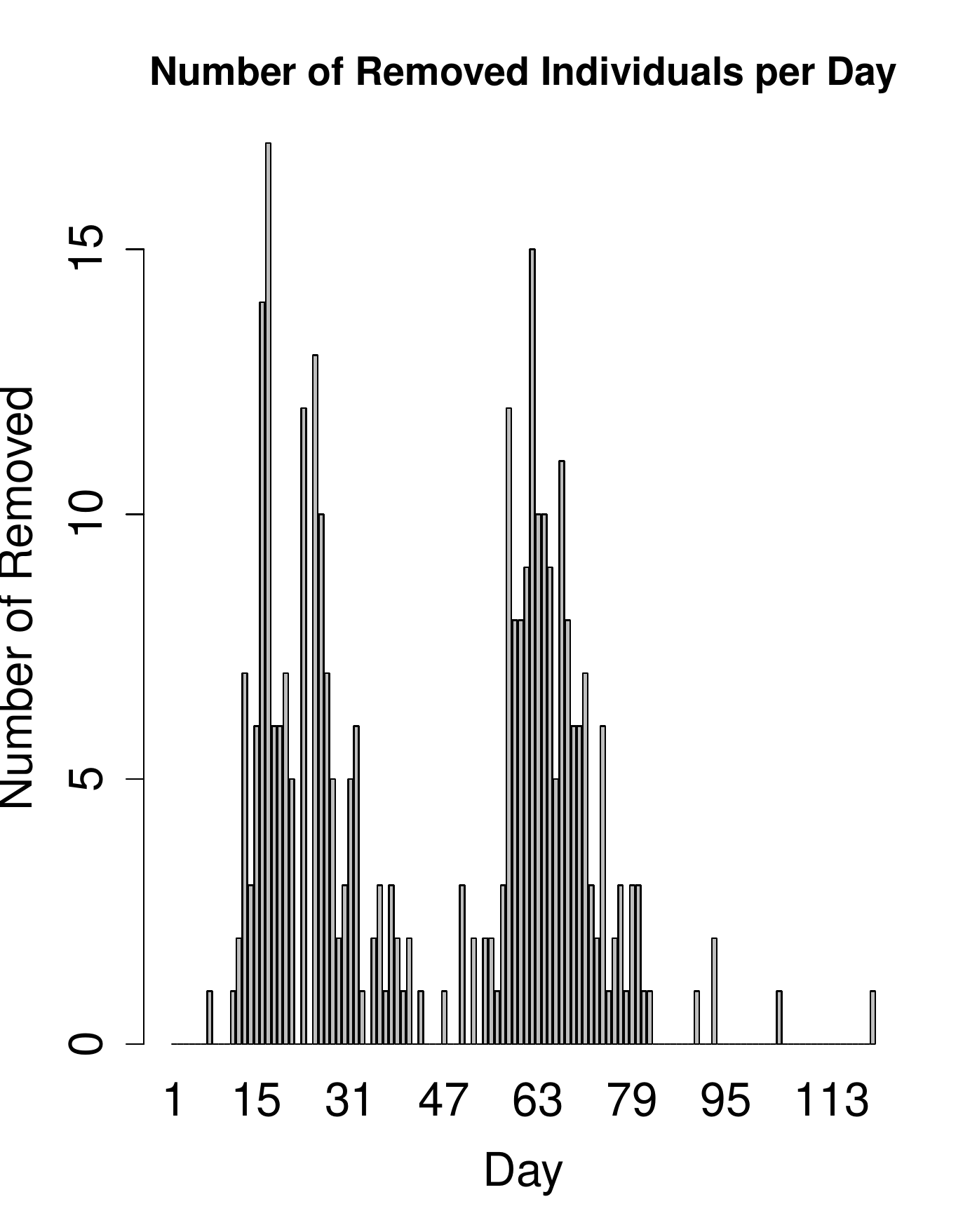}
\label{fig:subfigure1}}
\subfloat[True $\beta$ and ML estimates]{%
\includegraphics[height=7cm,width=7cm]{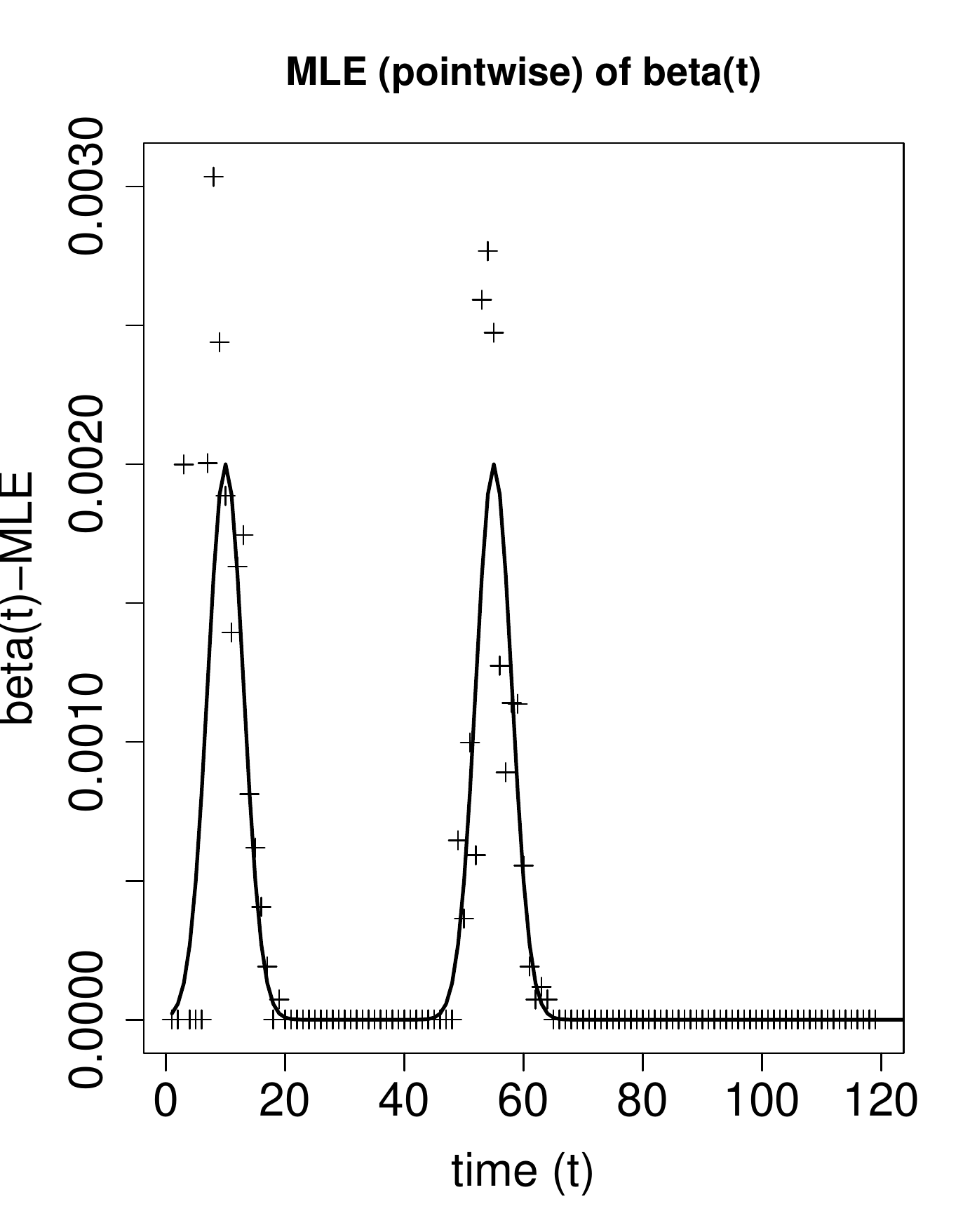}
\label{fig:subfigure2}}
\quad
\subfloat[Posterior $\beta$, infection times known]{%
\includegraphics[height=7cm,width=7cm]{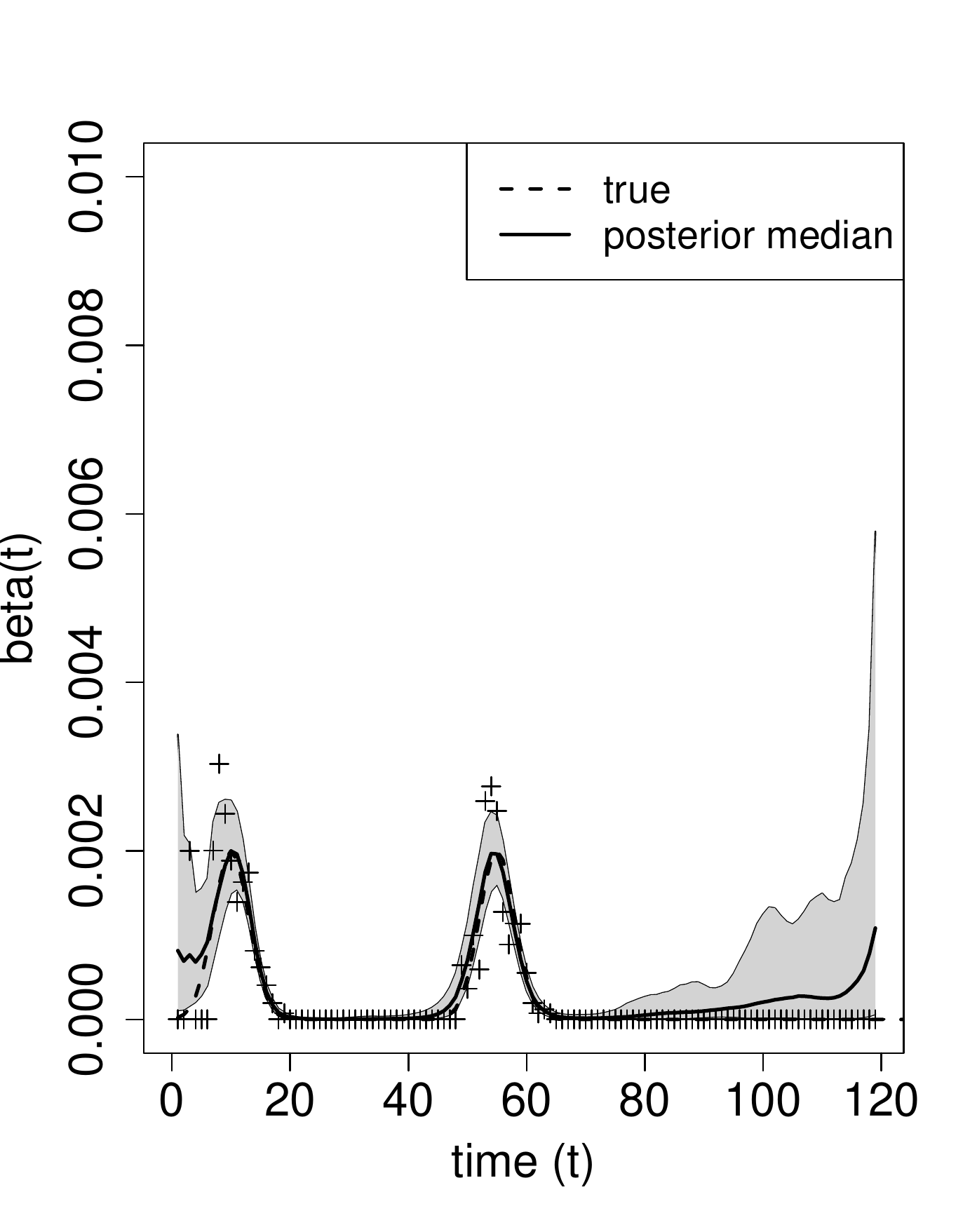}
\label{fig:subfigure3}}
\subfloat[Posterior $\beta$, infection times unknown]{%
\includegraphics[height=7cm,width=7cm]{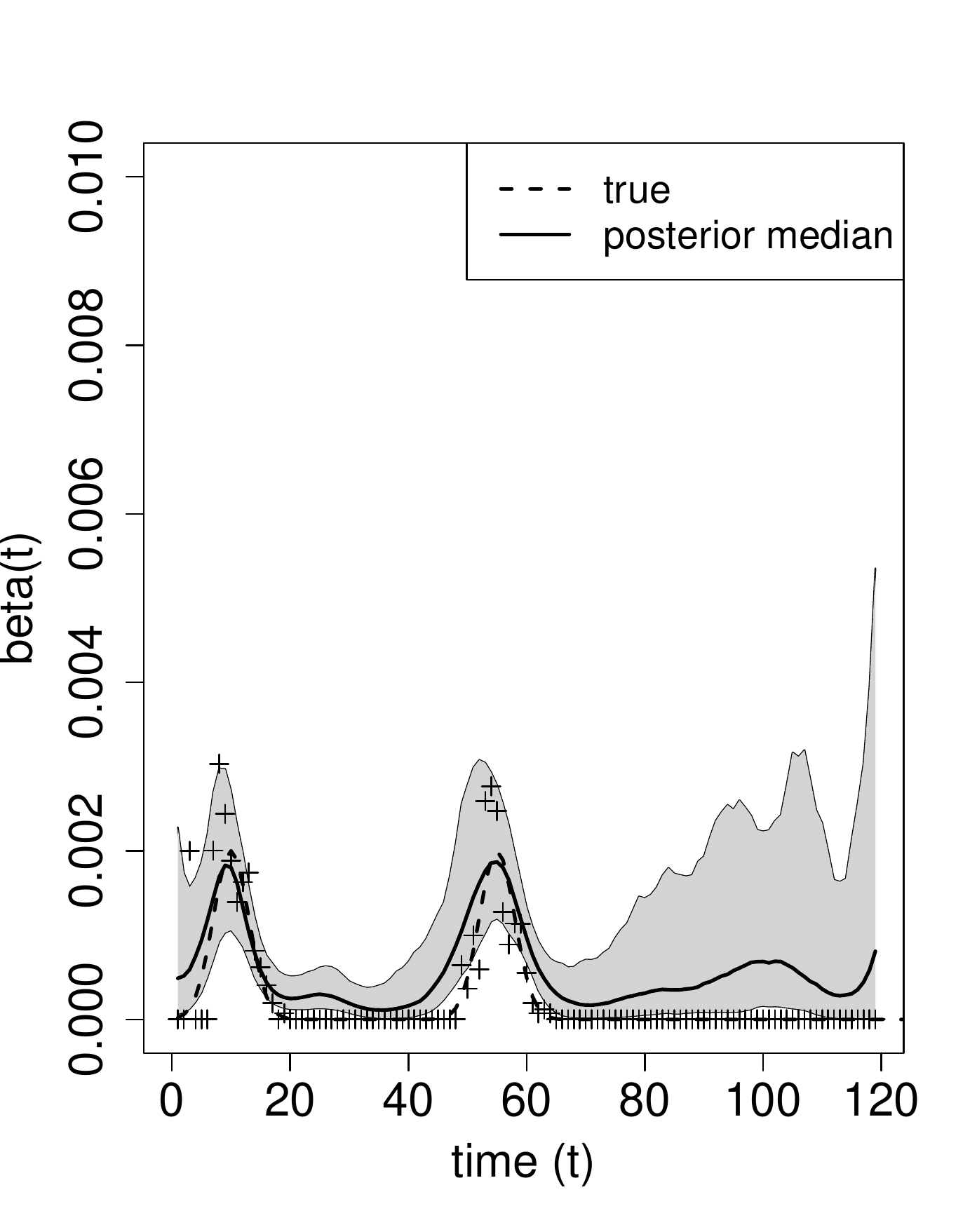}
\label{fig:subfigure4}}
\caption{Results for simulated data in Scenario 2 using $\beta(t) = (0.002) \exp\left\{ -(x-10)^2 /18 \right\} +
(0.002) \exp\left\{ - (x-55)^2 / 18 \right\}$, $\gamma = 0.1$, $N=500$, and GP prior covariance
function values $\omega = 8$, $l=5$. Plots (b) and (c) are
obtained using the known true infection times, and plot (d) only uses the removal data. The shaded regions in
(c) and (d) are 95\% posterior credible intervals for $\beta$. The ML estimates are shown as crosses in (c) and (d) for comparison.}
\label{seasonal_example}
\end{figure}

The results are illustrated in Figures \ref{exponential_example} and \ref{seasonal_example}. The methods appear to perform reasonably well in
practice. In all cases we see larger credible intervals for $\beta$ at the very start and at end of the outbreak. This is to be expected, since in
these times there are typically fewer infections from which to infer the value of $\beta$.\\

\begin{figure}[ht]
\centering
\subfloat[Symptom-appearances each day]{%
\includegraphics[height=7cm,width=7cm]{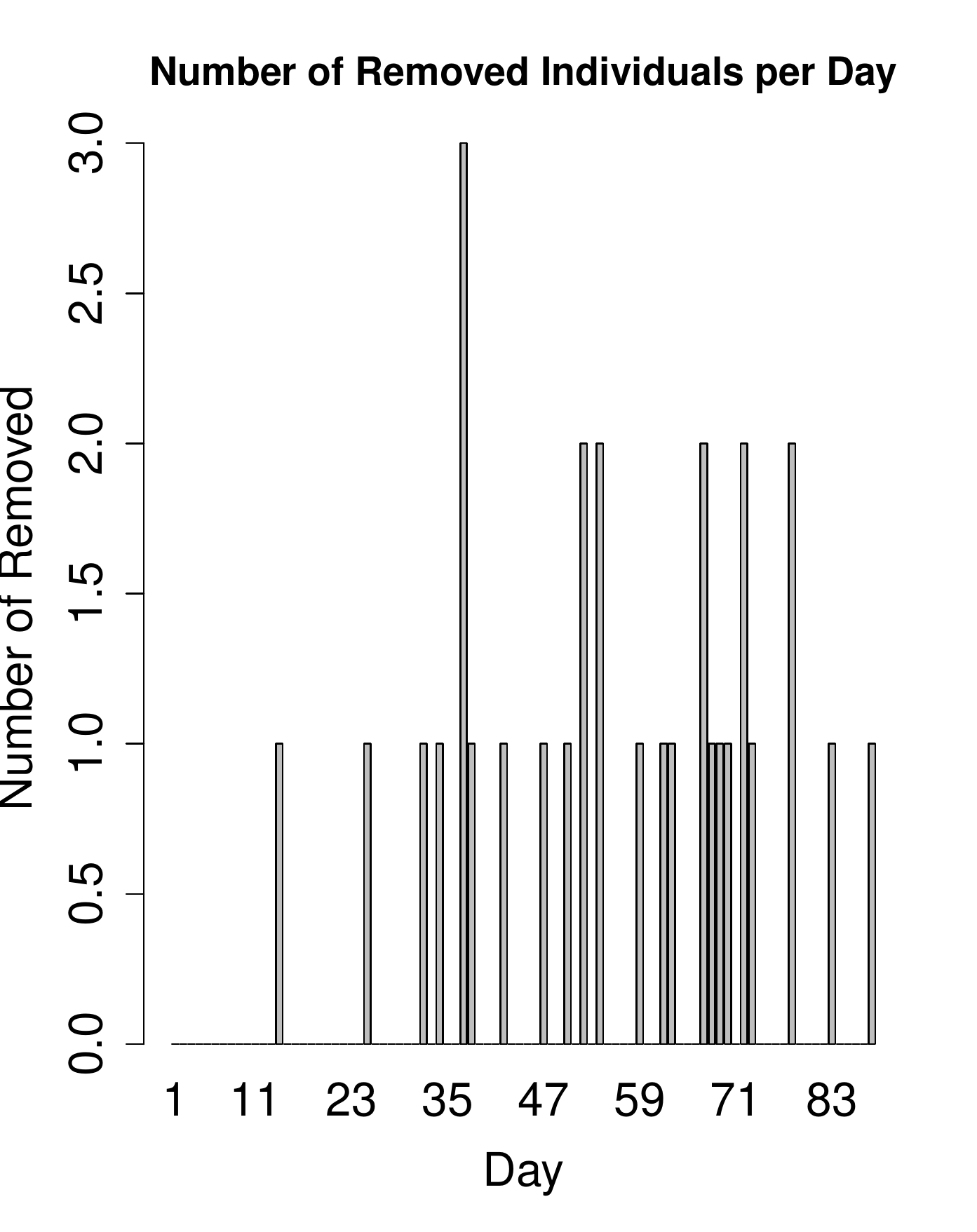}
\label{fig:poxsubfigure1}}
\subfloat[Posterior median of $\beta$]{%
\includegraphics[height=7cm,width=7cm]{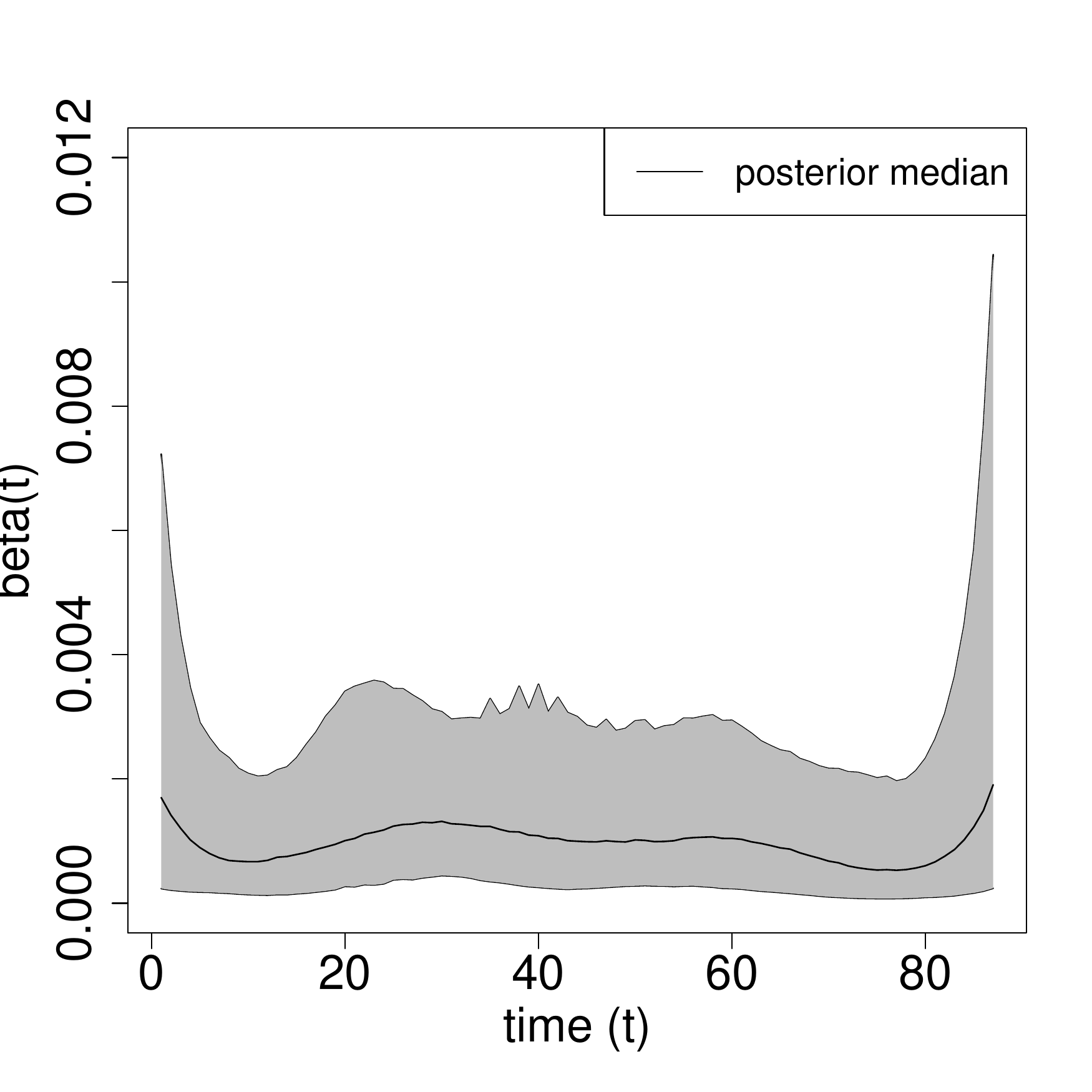}
\label{fig:poxsubfigure2}}
\quad
\subfloat[Posterior mean of $\beta$, continuous-time analysis]{%
\includegraphics[height=6cm,width=8cm,trim=0 100 0 100]{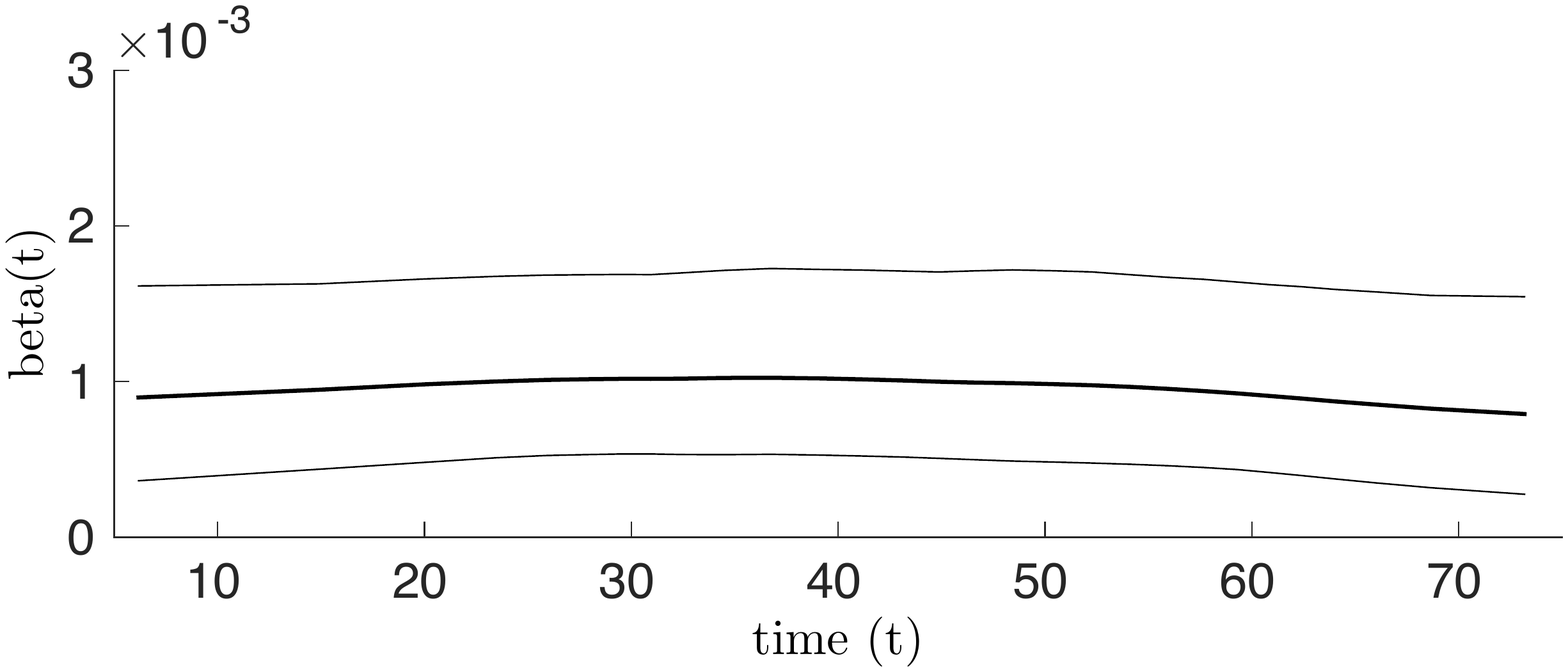}
\label{fig:poxsubfigure3}}

\caption{Results for the Abakaliki smallpox data (a,b). Here $N = 120$ and GP prior covariance
function values are $\omega = 5$, $l=14$. The shaded region in plot (b) shows the 95\% credible
interval for $\beta$. The figure in (c) is taken from a continuous-time analysis described in
\cite{xu15} and shows the posterior mean and 95\% credible interval for $\beta$.}
\label{smallpox_example}
\end{figure}

{\em Smallpox data} Our final example uses a classical data set collected during an outbreak of smallpox in the Nigerian
town of Abakaliki in 1967 \citep{thompsonfoege68}. These data have been considered by numerous authors, almost always to illustrate new
statistical methodology, and are usually taken to consist of the symptom-appearance times of 30 individuals among
a homogeneously-mixing susceptible population of 120 individuals. More extensive analyses of the full data set,
which includes information on population structure, vaccination and other aspects, can be found in \cite{eichner-dietz}
and \cite{stock17}. The results are shown in Figure \ref{smallpox_example}, the key finding of which is that there is
no evidence to suggest any material variation in the infection rate during the outbreak. For comparison with a continuous-time
analysis, we include a figure from \cite{xu15} which uses the methods described in Section \ref{subsec:GP for cts time}, so that
the infection rate has a transformed GP prior. It is clear that this approach gives similar results to our discrete-time
approach.

\section{Concluding comments}
\label{sec:conclusions}	
We have briefly described recently-developed methods for nonparametric Bayesian modelling and inference for epidemic models, specifically
focussing on infection or incidence rate functions in SIR models. The methods themselves can
easily be extended to other situations, including epidemic models in which there are several types of individuals
with potentially different infectivity of susceptibility characteristics, or more complex models that include
features such as latent periods or more realistic population structure. Some of these extensions are described in
\cite{xu15} and \cite{xuetal16}.

Other aspects of epidemic models could also be modelled nonparametrically. \cite{BoysGiles07} essentially do this
by replacing the removal rate $\gamma Y(t)$ by a more flexible time-varying function modelled by a step-function.
One motivation for doing this is to investigate the usual assumption that infectious periods are identically
distributed. Another possibility is to model population structure nonparametrically. In general this appears to
be a challenging problem, since it involves defining a suitable prior for population structure, which could itself be
described by a network or random graph model. An approach involving the so-called Indian Buffet Process (IBP)
is described in \cite{ford}, in which the IBP is used as a model for a bipartite graph on which an epidemic
spreads. However, the resulting inference problem generates many computational challenges.

\section*{Acknowledgements}
We thank two reviewers and the Associate Editor for helpful comments that have improved the manuscript.

\newpage


\begin{thebibliography} {99}

\bibitem[Adams {\em et al.} (2006)]{adams-murray06}
\textsc{Adams, R. P., Murray, I., and MacKay, D. J. C.} (2006).	
The Gaussian process density sampler. In
\emph{Advances in Neural Information Processing Systems 21},
Eds. Koller, D., Schuurmans, D., Bengio, Y. and Bottou, L.,
pp. 9--16.

\bibitem[Adams {\em et al.} (2009)]{adams-murray09}
\textsc{Adams, R. P., Murray, I., and MacKay, D. J. C.} (2009).
\emph{Tractable nonparametric Bayesian inference in Poisson processes with Gaussian process intensities.}
New York: ACM Press.
			
\bibitem[Andersen \em{et al.} (1993)]{andersen1993}
\textsc{Andersen, P. K., Borgan, \O, Gill, R. D. and Keiding, N.} (1993).
\emph{Statistical Models based on Counting Processes}.
New York: Springer.

\bibitem[Andersson and Britton(2000)]{andbritt2000}
\textsc{Andersson, H. and Britton, T.} (2000).
\emph{Stochastic Epidemic Models and their Statistical Analysis.}
Lecture Notes in Statistics, 151.
New York: Springer.


\bibitem[Bailey(1975)Bailey]{bailey}
\textsc{Bailey, N. T. J.} (1975).
\emph{The Mathematical Theory of Infectious Diseases and its Applications, 2nd ed.}
London: Griffin.
	
\bibitem[Becker(1989)]{becker89}
\textsc{Becker, N. G.} (1989).
\emph{Analysis of Infectious Disease Data.}
London: Chapman and Hall.
		
\bibitem[Becker and Yip(1989)Becker and Yip]{becker-yip}
\textsc{Becker, N. G. and Yip, P. S. F.} (1989).
Analysis of variation in an infection rate.
\emph{Australian Journal of Statistics}
\textbf{31},
42--52.

\bibitem[Boys and Giles, 2007]{BoysGiles07}
\textsc{Boys, R. J. and Giles, P. R.} (2007).
\newblock{Bayesian inference for stochastic epidemic models with time-inhomogeneous
removal rates}.
\newblock{\em Journal of Mathematical Biology} {\bf 55}, 223--247.
			
\bibitem[Chen {\em et al.}(2008)]{chen}
\textsc{Chen, F., Huggins, R. M., Yip, P.S. and Lam, K. F.} (2008).
Nonparametric estimation of multiplicative counting process intensity functions with an application to the Beijing SARS epidemic.
\emph{Communications in Statistics: Theory and Methods}
\textbf{37},
294--306.
	
\bibitem[Eichner and Dietz(2003)Eichner and Dietz]{eichner-dietz}
\textsc{Eichner, M. and Dietz, K.} (2003).
Transmission potential of smallpox: estimates based on detailed data from an outbreak.
\emph{American Journal of Epidemiology}
\textbf{158},
110--117.
			
\bibitem[Ford(2014)]{ford}
\textsc{Ford, A. P.} (2014)
Epidemic models and MCMC inference.
PhD Thesis, University of Warwick.
{{\tt http://webcat.warwick.ac.uk/record=b2752139\textasciitilde S1}} .


\bibitem[Knock and Kypraios (2016)]{knock}
\textsc{Knock, E. S. and Kypraios, T.} (2016).
Bayesian non-parametric inference for infectious disease data.
{{\tt https://arxiv.org/abs/1411.2624}}.

			
\bibitem[Lau and Yip(2008)Lau and Yip]{lau-yip}
\textsc{Lau, E. H. Y. and Yip, P. S. F.} (2008).
Estimating the basic reproductive number in the general epidemic model with an unknown initial number of susceptible individuals.
\emph{Scandinavian Journal of Statistics}
\textbf{35},
650--663.
			
\bibitem[Lekone and Finkenstadt(2006)]{Lekone}
\textsc{Lekone, P. E. and Finkenst\"{a}dt, B. F.} (2006).
Statistical inference in a stochastic SEIR model with control intervention: Ebola
as a case study.
\emph{Biometrics}
\textbf{62},
1170--1177.

\bibitem[McKinley {\em et al.}(2009)]{cook09}
\textsc{McKinley, T. J., Cook, A. R. and Deardon, R.} (2009).
Inference in epidemic models without likelihoods,
{\em The International Journal of Biostatistics} {\bf 5(1)}.

\bibitem[O'Neill and Roberts(1999)[O'Neill and Roberts]{oneill-roberts}
\textsc{O'Neill, P. D. and Roberts, G. O.} (1999).
Bayesian inference for partially observed stochastic epidemics.
\emph{Journal of the Royal Statistical Society Series A}
\textbf{162},
121--129.
			
\bibitem[Pollicott \emph{et al.}(2012)%
Pollicott, Wang and Weiss]{pollicott12}
\textsc{Pollicott, M. Wang, H. and Weiss, H.} (2012).
Extracting the time-dependent transmission rate from infection data via solution of an inverse ODE problem.
\emph{Journal of Biological Dynamics}
\textbf{6},
509--523.


\bibitem[Rasmussen and Williams(2006)%
Rasmussen and Williams]{rasmussen06}
\textsc{Rasmussen, C. E. and Williams, C. K. I.} (2006).
\emph{Gaussian Processes for Machine Learning.}
Cambridge, Massachusetts: MIT Press.

\bibitem[Smirnova and Tuncer(2014)Smirnova and Tuncer]{smirnova-tuncer}
\textsc{Smirnova, A. and Tuncer, N.} (2014).
Estimating time-dependent transmission rate of avian influenza via stable numerical algorithm.
\emph{Journal of Inverse and Ill-Posed Problems}
\textbf{22}(1),
31--62.


\bibitem[Stockdale {\em et al.}(2017)]{stock17}
\textsc{Stockdale, J. E., Kypraios, T. and O'Neill, P. D.} (2017).
Modelling and Bayesian analysis of the Abakaliki smallpox data.
{\em Epidemics}, in press.


\bibitem[Thompson and Foege, 1968]{thompsonfoege68}
\textsc{Thompson, D. and Foege, W.} (1968).
\newblock{Faith Tabernacle smallpox epidemic. Abakaliki, Nigeria.}
\newblock{World Health Organization.}
				
\bibitem[Xu, 2015]{xu15}
\textsc{Xu, X.} (2015).
\newblock{Bayesian Nonparametric Inference for Stochastic Epidemic Models.}
\newblock{Ph.D. thesis, University of Nottingham.}
{\tt http://eprints.nottingham.ac.uk/29170/}

\bibitem[Xu {\em et al.} (2016)]{xuetal16}
\textsc{Xu, X., Kypraios, T. and O'Neill, P. D.} (2016)
\newblock{Bayesian non-parametric inference for stochastic epidemic models using
Gaussian processes.}
\em{Biostatistics}
\textbf{17},
619--633.



				
\end{thebibliography}
\end{document}